\documentclass{article}

\usepackage[english]{babel}

\usepackage[letterpaper,top=2cm,bottom=2cm,left=3cm,right=3cm,marginparwidth=1.75cm]{geometry}

\usepackage{amsmath}
\usepackage{graphicx}
\graphicspath{{images/}}
\usepackage[colorlinks=true, allcolors=blue]{hyperref}
\usepackage{authblk}
\title{XLO-II, a high-repetition rate x-ray laser oscillator}
\author[1]{C. Pellegrini}
\author[1]{A. Halavanau}
\author[2]{A. Benediktovitch}
\author[3]{U. Bergmann}
\affil[1]{SLAC National Accelerator Laboratory, Menlo Park, California}
\affil[2]{Center for Free Electron Laser Science, DESY, Hamburg, Germany}
\affil[3]{University of Wisconsin-Madison, Madison, Wisconsin}
\begin{document}
\maketitle

\begin{abstract}
In a recent paper we proposed to build an x-ray laser oscillator (XLO) in the 6-10 keV range providing intense, stable, transform-limited, x-ray pulses based on population inversion driven by an x-ray  pulse train generated  by an x-ray free-electron laser (XFEL) operated at a repetition rate of about 100 Hz. Here we present an analysis of recent experimental results on x-ray lasing with population inversion, damage caused by the pump on the lasing medium, and optical cavities, together with theoretical/numerical simulations. Our findings suggest that it is possible to build and operate a second-generation x-ray laser oscillator, XLO-II,  operating at up to 125 kHz repetition rate. XLO-II will be pumped by 6-10 keV x-ray SASE pulses, generated by the new LCLS-II-HE XFEL now under construction at SLAC National Accelerator Laboratory, utilizing a CW superconducting linac and capable of running at 1 MHz repetition rate. XLO-II will generate transform-limited, coherent x-ray pulses with an average power in the tens of mW range. It will open new experimental capabilities, for instance in fields like imaging, interferometry, and quantum x-ray optics. The main characteristics of XLO-II and its main components, including the optical cavity, will be discussed.
\end{abstract}

\section{Introduction}

\begin{figure}[h!]
    \centering
    \includegraphics[width=1\linewidth]{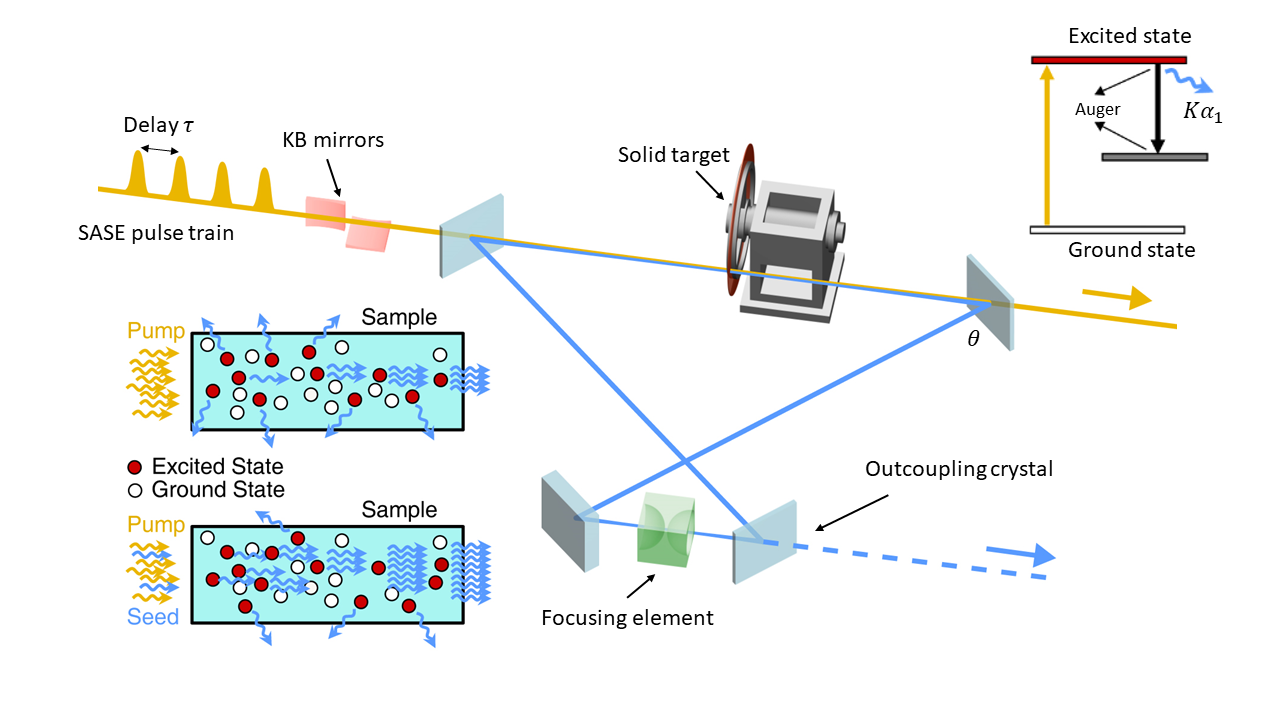}
    \caption{Schematics of XLO-II (not to scale).  A 9~keV SASE pump pulse impinges on a fast-moving copper foil gain medium. The resulting stimulated x-ray emission at 8.048 keV is recirculated in the Bragg bow-tie cavity and overlapped with the consecutive pump pulse arriving 1 $\mu$s later. This is repeated until saturation is reached. The copper foil must move fast enough to present an undisturbed, fresh copper volume to each consecutive pump pulse.}
    \label{fig:layout}
\end{figure}

A new x-ray free-electron laser (XFEL), LCLS-II-HE (Linac Coherent Light Source II High Energy)\cite{Raubenheimer:FLS2018-MOP1WA02}, driven by an 8 GeV, CW superconducting linac, is under construction at SLAC National Accelerator Laboratory (SLAC).  The beam will feed two undulators (one in the soft x-ray (SXR) regime and one in the hard x-ray (HXR) regime) to generate high intensity self-amplified spontaneous emission (SASE) pulses with repetition rates up to 1 MHz. The hard x-ray (HXR) undulator will produce x-ray pulses with a photon energy up to 15 keV. In this paper we discuss the feasibility of using SASE pulses from the LCLS-II-HE HXR undulator to pump a population inversion-based x-ray laser oscillator, that we call XLO-II, operating in the 6-10 keV photon energy range. The schematics of XLO-II is shown in Figure  \ref{fig:layout}. The basic concept of XLO-II is similar to that of XLO \cite{Alex1}. A population-inversion gain medium pumped with a focused XFEL pulse is placed in the center of one of the two parallel arms of a back-scattering Bragg crystal cavity. Additional x-ray focusing optics is placed in the cavity to ensure good spatial overlap with the pump pulse. The main conceptual difference between XLO-II and XLO is the pump pulse time separation. While XLO uses as a pump a SASE XFEL pulse train generated at 120 Hz repetition rate by the SLAC copper linac with a temporal pulse spacing of $\approx 35$ ns, matching the $\approx 10$ m cavity length, XLO-II will use a larger pulse spacing of 1000 ns (corresponding to 1 MHz repetition rate), and, depending on the cavity length, each pulse will be recycled between 10 and 30 times through the cavity before the next pulse arrives.

In the case that we study and present here the oscillator saturates in four pulses and XLO-II will operate at repetition rates of 125 kHz, providing stable x-ray pulses with more than $\approx 10^{8}$ photons per pulse, each pulse being fully coherent and transform-limited. XLO-II has similar intensity and coherence properties per pulse as XLO, but with three orders of magnitude increase in repetition rate, and thus in average power and brightness. Generating stable, transform-limited, fully coherent pulses, with a spectral bandwidth $\Delta$E/E of the order of $10^{-5}$ and up to 125 kHz repetition rate, will open new research fields beyond the reach of the existing x-ray FELs.  In particular, it will allow new research in x-ray quantum optics, interferometry, coherent imaging and other areas, where a well-defined phase of the x-ray pulse used to probe matter and/or intensity stability, is critical. 

The XLO-II components, performance, and feasibility are based on recent experimental and theoretical/simulation results that are discussed in the following sections.
The first important development has been the improved understanding and measurement of the parameters that generate seeded stimulated x-ray emission based on population inversion of 1s core hole states created by a SASE pump pulse\cite{Rohringer2012, Yoneda, PhysRevLett.120.133203, PhysRevLett.125.037404, doi:10.1073/pnas.2119616119}. We recently demonstrated that inner-shell x-ray lasers at 6-10 keV can be operated using highly focused x-ray FEL pump pulses in the 10-50 $\mu$J energy range, corresponding to about $10^{18}$-$10^{19}$ W/$cm^2$, when a weak seed pulse is employed \cite{osti_1972766, Halavanau:22, Alex2, 10.1063/5.0168125}. The second important development has been a quantitative understanding of the damage induced by the pump pulses in a solid copper gain medium as a function of pump pulse energy/intensity. Lower pump pulse energies cause smaller craters in the gain medium, thus making it easier to have a fresh part of the medium ready for exposure to a new pump pulse \cite{10.1063/5.0168125, Manwani:2022rmr}. The third important development has been the progress in the development of low-loss x-ray optical cavities using Bragg reflections in Si or diamond crystals, in rectangular or bow-tie configurations. It has been recently demonstrated experimentally that a diamond-based optical Bragg cavity with very low losses of order of 1\% per crystal at 9 keV \cite{Margraf2023} can be operated with an x-ray beam cycling through the cavity many times. We will discuss these three important developments in the following sections and show how they lead to the feasibility of XLO-II.

To help analyze the characteristics of XLO-II we use a new numerical simulation code \cite{benediktovitch2023stochastic},  based on a 3D model of the lasing process in a copper gain medium, using LCLS-II-HE SASE x-ray FEL pump pulses to generate a population inversion. The results of the numerical code agree with recent experimental results from our work on XLO and other projects. 
In what follows we first discuss the characteristics of the x-ray pump pulses generated by LCLS-II HE, we then describe the recent measurements on population inversion x-ray lasers, the results on optical cavity losses and on the gain medium damage induced by the pump pulses. We then give the results of numerical simulations of XLO-II, provide the oscillator's initial design, and summarize our findings, discussing its characteristics.

\section{LCLS-II HE photon pulses characteristics }

The estimated performance of LCLS-II HE, based on numerical simulations,  is reported in  Ref. \cite{osti_1884842}. The performance is evaluated for three cases: a) 100 pC electron bunch with a time duration of 40 fs; b) 20 pC electron bunch with a time duration of 20 fs; c) an advanced low emittance injector. We are mainly interested in case b), that, according to our results on lasing with seeding discussed later, will provide enough power density in the gain medium to pump the oscillator.
As shown in Ref. \cite{Raubenheimer:FLS2018-MOP1WA02} and \cite{osti_1884842} at 9 keV the energy per pulse in SASE mode is about 300 $\mu$J when operating in the 20 pC case. In this case the system can deliver the pulses at 1 MHz repetition rate. This pulse energy at 1 MHz repetition rate puts beam dump and the optical instrumentation following the undulator to their thermal limit at 300 W average. Since we want to use part of the undulator for generating a weak seed pulse at 8.048 keV, we use in our simulation 200 $\mu$J pump pulses, to stay below the thermal limit. Then, the pump pulse energy delivered to the gain medium at 9 keV is 100 $\mu$J at the focal point, assuming a 50\% reduction due to losses in the KB nanofocusing system. It has been shown in recent experimental work \cite{osti_1972766, Alex2} that the use of a weak seed pulse can dramatically reduce the pump energy needed for lasing. This important point for the feasibility of XLO-II is discussed now.

\section{Recent results on population inversion x-ray lasers}
\label{recent-results}
\begin{figure}
    \centering
    \includegraphics[width=0.65\linewidth]{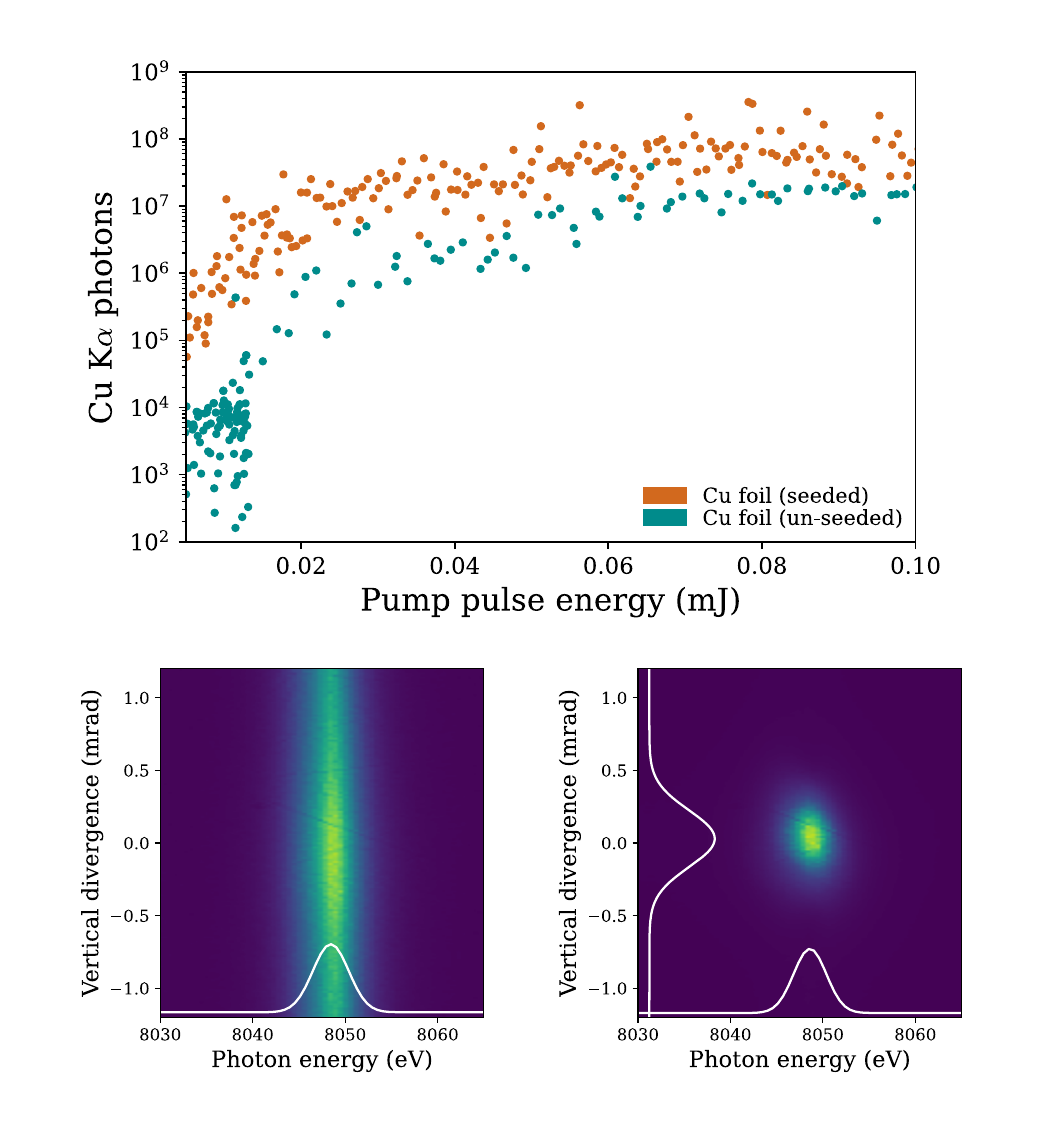}
    \caption{Top: Number of $K\alpha_1$ copper photons as a function of pump pulse intensity for undulator seeded and unseeded cases. The data show about $10^3$ background photons and about $10^5$ seed photons from 4 LCLS HXUs. In the seeded case about $10^8$ photons are obtained with a pump pulse energy of the order of 50 $\mu$J. Bottom row:  Measured angular/energy distribution of $K\alpha_1$ photons in the vertical direction after selection by a Si (220) analyzer crystal for the ASE (left) and the undulator seeded (right) cases; color scale is arbitrary. The angular width in the seeded case is largely defined by the KB mirror focusing of the undulator radiation, about 0.5 mrad FWHM; in the ASE case, by the gain medium radius to length ratio, $>10$ mrad.}
    \label{fig:XLO II gain curves}
\end{figure}
Recent results in population inversion lasing in solid copper are shown in Fig. \ref{fig:XLO II gain curves}, where the number of photons generated at the $K\alpha_1$ line is plotted as a function of the 9 keV pump pulse intensity in a solid copper foil of 25 $\mu$m thickness. Pump pulses of an estimated 10 fs pulse length are focused to a size of $\approx$100 nm diameter FWHM using the nanofocus KB optics \cite{seaberg2019nanofocus} at the CXI instrument at LCLS.
For our present evaluations, we assume a Gaussian x-ray pulse with a radial intensity distribution given, for total pulse energy $E_{FEL}$, by 
$I=\frac{2 E_{FEL}}{\pi w_0^2}\exp\{-2 r^2/w_0^2\}$.
The peak energy density is $2E_{FEL}/\pi w_0^2 $. Assuming the FWHM to be 120 nm, as measured in our XLO experiment using CXI, we have $1/e^2$ width of $2 w_0=203$ nm. The Rayleigh range and the angular divergence are $Z_R=\pi w_0^2/\lambda_p$ = 200 $\mu$m, $\theta=2w_0/Z_R$= 1 mrad. The effective area is $A=1.6\times 10^{-10} cm^2$. The peak energy density at the nano-focus is $E_d =6 \times 10^9 E_{FEL} J/cm^2$. 
The power density for this condition is $1.6\times10^{19}$ W/cm$^2$ at 50 $\mu$J and the corresponding energy density $1.6\times10^5$ $J/cm^2$ .
From the results shown in Figure \ref{fig:XLO II gain curves} we can evaluate the power and energy density needed to obtain stimulated x-ray emission in the copper gain medium and the number of photons we obtain for a given pump volume, with and without seeding. Figure 2 shows the energy/angular distribution of the stimulated x-ray emission without seed (left) and with seed (right) measured with a flat Si(2,2,0) analyzer followed by a 2D position sensitive detector (see e.g. Ref. \cite{PhysRevLett.125.037404}). Seeded stimulated x-ray emission has a much smaller angular divergence (500 $\mu$rad) as compared to ASE. This is an important result for estimating the Bragg cavity losses.

\begin{figure}[h!]
    \centering
    \includegraphics[width=0.5\linewidth]{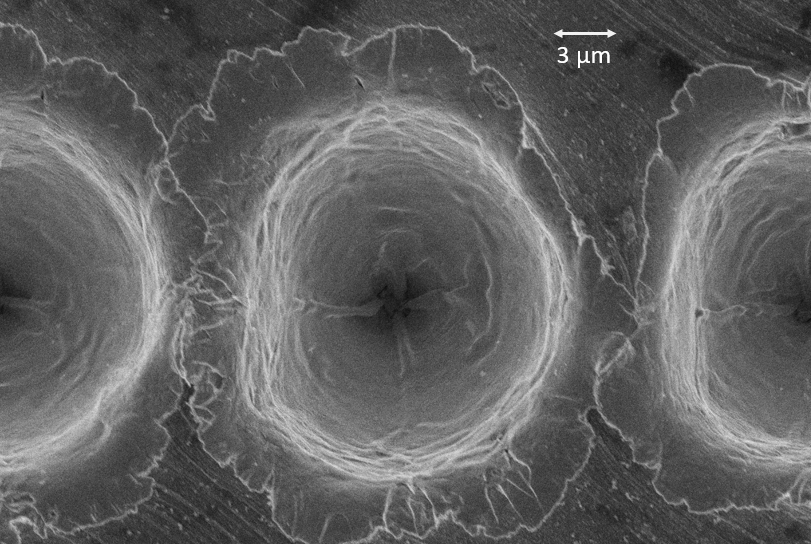}
    \caption{Scanning electron microscope image of a conical crater created by a 40 $\mu$J pulse in a 25 $\mu$m thickness copper foil.}
    \label{fig:XLO II copper damage}
\end{figure}

Another critical element needed to design and operate XLO-II is the copper gain medium. Here, information on the radial size of the damage produced by the pump pulse is needed to estimate the required speed and gain medium size for continuing XLO-II operations. After previously considering a liquid jet system using a copper solution \cite{Alex1}, we performed experiments where we found that the rapid sample replacement is very difficult in a liquid jet and that the lower copper density as compared to a solid copper foil limits the gain and the minimum pump pulse energy. We, therefore, moved to a new spinning disk system using a 25 $\mu$m thick copper foil and studied the effect on the crater size as a function of pump pulse energy \cite{10.1063/5.0168125, Manwani:2022rmr}. Figure \ref{fig:XLO II copper damage} shows the case of the damage produced by a 50 $\mu$J pump pulse \cite{Alex2}. The damaged ablation area of the copper target has a diameter of 14 $\mu$m, reduced to 10 $\mu$J at 40 $\mu$J.
 
The results of damage can be summarized in two empirical formulas, one giving the radial size of the ablation damage, $r_a=8.4 \ln E_p+36.6$, the other the radial size of the through hole created by the pump pulse, $r_h=28.6 E_p-2.2$ [applies for pulse energies above 0.1 mJ], where $E_p$ is in mJ and the size is in $\mu$m. 
These results allow us to establish the speed at which the gain medium, in this case copper, must be refreshed so that the next pump pulse will hit a fresh, unperturbed part of the gain medium.

The third important result has been the measurement of the losses of a 9.831 keV photon pulse circulating in a diamond crystal Bragg optical cavity at LCLS \cite{Margraf2023}. The losses have been measured over many revolutions for a 14 m long rectangular cavity, injecting a SASE pulse from the hard x-ray undulator into it. From the results, shown in Ref. \cite{Margraf2023},  the authors confirm that, after the initial revolutions  the loss per Bragg reflection is less than 1\% , in accordance with theoretical estimates.
The large initial losses are due to angular and energy filtering of the SASE pulse, with large energy and angular spread, by the Bragg crystals. Subsequent small losses are due to the crystal reflectivity and cavity alignment. For instance, in the time interval between 0.5 to 1.2 $\mu$s, after about 14 revolutions, the loss is a factor of ten, as shown in Ref. \cite{Margraf2023}.

The initial loss has also been partly measured in the XLO bow tie cavity using Si(4,4,4) crystals at 8.048 keV \cite{Alex2}.
In this case, the most important quantity measured is the intensity loss of a SASE pulse after reflections by the Si crystals. The measured intensity ratio before and after the crystal is then compared with the theoretical results. For the first crystal after the gain medium, the measured intensity ratio is 0.65 \%, and in good agreement with the calculated value of 0.7 \%. As mentioned before, the large loss is due to filtering in angle and energy and is in good agreement with theoretical models. The losses on the following crystal are also in agreement with the Si crystal reflectivity.

\section{Conceptual design of XLO-II}

\begin{figure}[h]
    \centering
\includegraphics[width=0.8\linewidth]{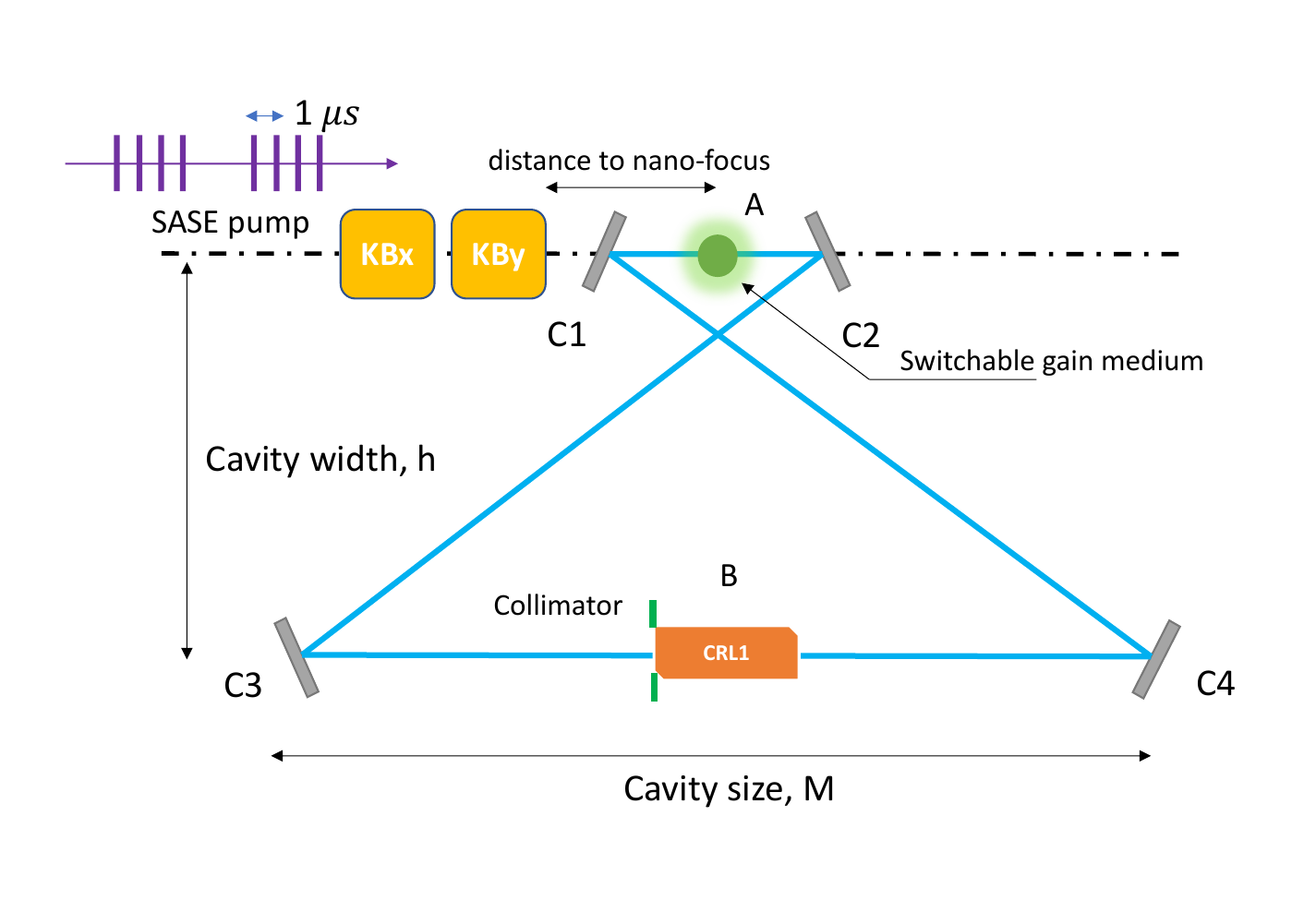}
    \caption{XLO-II schematic, considering the case when 4 pump pulses are needed to reach saturation. C1 - C4 are diamond crystals, reflecting x-rays at the Bragg angle. CRL1 and CRL2 are refractive lenses to focus the x-rays on the gain medium in point A, which is also the KB mirrors focal point. From A to B is half cavity length, $\L_C/2$. In a bi-concave focusing cavity, imaging point A back to itself, the required CRL focal length is $f=\L_C/4$. Other options for focusing are discussed in the text.}
    \label{fig:XLO II schematic}
\end{figure}

 We now use the results of the previous discussion to estimate the feasibility and performance of XLO-II. To do this we consider the design of the optical cavity and later that of the gain medium system. The XLO-II schematic is shown in Figure \ref{fig:XLO II schematic}.

LCLS-II-HE operates at 1 MHz, the time separation between pump pulses is 1 $\mu$s or longer. The simplest solution to synchronize the pump pulses with those circulating in the cavity is to build a 299.8 m long cavity, with a 1 $\mu$s circulation time. However, a long cavity is technically more challenging to build and operate, as it is more susceptible to spatial, angular, and temperature instabilities. To overcome these challenges, we can take advantage of the low losses of diamond Bragg crystals used as reflectors (compared to previously considered silicon crystals), that allow multiple circulations of the amplified pulse in the cavity before seeding the amplification when the next pump pulse arrives. Recently more than 60 circulations have been observed in a diamond Bragg cavity with 86\% transmission for each circulation \cite{Margraf2023}. 

To design the system, we consider the case studied above with our numerical simulations using four pump pulses to amplify to saturation followed by out-coupling. In the discussion of the evolution of the XLO pulse circulating in the cavity while being amplified in the gain medium, we differentiate between the first time the pulse circulates the cavity and the subsequent times. In the first amplification, we start with a 40 $\mu$J, 9 keV pump pulse generated by the main section of the undulator to create the population inversion, and a few $\mu$J seed pulse generated from a small section of the undulator tuned to the Mn K$\alpha$ line at 8.048 keV. Based on the results shown in Figure \ref{fig:XLO II gain curves}, a stimulated emission pulse at 8.048 keV with about $10^8$ photons can be generated. When using a seed pulse, we observe an angular divergence of about 500 $\mu$rad FWHM and an energy bandwidth of about 4 eV FWHM. Only part of these photons will fall within the first Bragg diamond crystal Darwin width, where the reflectivity is near 98\% (the losses of the diamond crystals at 8.048 keV are assumed to be about 2\% per crystal). Using the previous numbers for the angular and energy spread of the amplified pulse and the characteristics of a diamond (4,0,0) crystal given in Table \ref{tab:crystals} we estimate a reduction in the number of photons after being reflected by the first crystal by a factor $2\cdot10^{-2}$, and a factor of $3\cdot10^{-4}$ reduction after four crystals, thus reducing the number of photons propagating through the rest of the cavity to about $10^5$. The difference between this value and that reported in Ref. \cite{Alex2} and mentioned above, is due to the use of a diamond crystal, with smaller Darwin width, instead of a silicon crystal.

These photons must be circulated in the cavity and focused to a spot size equal to that of the pump pulse, with a FWHM of 120 nm. 
To focus the x-ray pulse on the pumping volume, we need to introduce focusing elements in the cavity. There are several ways to do this. Here we discuss one using a collimator to reduce the vertical angular spread to the same value the first crystal introduces in the horizontal plane, and CRL lenses. We will evaluate the photon losses due to the collimator and the CRL lenses and evaluate the associated photon losses. The collimator reduces the vertical angular distribution from 500 $\mu$rad to 14 $\mu$rad, the same of the horizontal direction, and introduces a cut of another factor of 20 in the number of amplified photons, to about $1.5\times10^4$ photons. This number must be further reduced by the losses in the 3 additional crystal and the losses in the CRL lenses. 
The choice of these focusing elements, and the additional losses they introduce, are discussed later and shown to be in the 10\%  range. Thus we estimate the number of photons that will arrive back to the pumped volume and serve as a seed to be about $10^4$. This number is large enough to generate about $10^8$ stimulated emission photons that, like the seed signal,  are contained in an angular aperture of 14 $\mu$rad and an energy width of 0.06 eV. In the next passage through the optical cavity the losses are now due only to the crystals and CRLs absorption and will be in the range of 10$\%$ per evolution.

To avoid the collimator loss associated to the reduction of the vertical angular spread it is possible to consider other system. One is focusing in the vertical plane by bending the diamond crystals around an horizontal axis and in the horizontal 
with cylindrical CRLs. Another is the use of glancing angle elliptical mirrors. These, and other alternative cases, will be discussed in future papers. It is important to notice that the case we discuss is sufficient to establish the feasibility of the proposed oscillator.

\subsection{The optical cavity focusing system}

The losses in the cavity following the first circulation will determine the number of cavity passes without amplification we can use to reduce the cavity length. 
We have already discussed the loss following the first amplification event, seeded by the undulator radiation, in the first cavity pass. In the following passes, before the next amplification event takes place, the pulse will match the angular and energy characteristics of the Bragg crystals, and the losses will only be due to the crystal and focusing system absorption. 
 
We now consider what can be accomplished using the diamond (4,0,0) crystals as mirrors and compound refractive lenses (CRLs) to focus the beam.
The cavity in the LCLS 14-m long cavity ringdown experiment described before \cite{Margraf2023} has rather weak focusing. In our case we must consider the need to focus the amplified pulse generated by the gain medium to the same size of the volume where we create population inversion after one or more round trips in the cavity. The population inversion spot size is about one hundred nm FWHM horizontally and vertically. We must also evaluate the effect that the focusing lenses has on losses. As an example for our discussion we consider a cavity length of 30 m. The results can be easily generalized to other sizes. The cavity is shown in Figure \ref{fig:XLO II schematic}. Point A is the KB mirrors focal point and the gain medium location. From A to B is half cavity length, $\L_C/2$. For a bi-concave CRL located at point B, imaging point A on itself requires a focal length $f=\L_C/4$.
In this case we want to use a bifocal lens and a collimator to cut the vertical angular spread to the same value defined by the Darwin width in the horizontal plane, $14 \mu$rad for diamond (400). The CRL can be built with beryllium to minimize the intensity losses. The beryllium refractive index at 8.048 keV is $n = 1-\delta+i\beta$ with $\delta= 5.318 \times 10^{-6}$ \cite{Serebrennikov:ay5492}. The focal length of a single bi-concave lens is $f=R/2\delta$. For $N$ lenses we have $f=R/2N\delta$.

If we want $f = 7.5$ m we need $R/N=2f\delta=7.7×10^{-5}$ m. For $N=2$ we have $R\approx150$ $\mu$m. 

 \begin{figure}[h!]
    \centering  \includegraphics[width=0.8\linewidth]{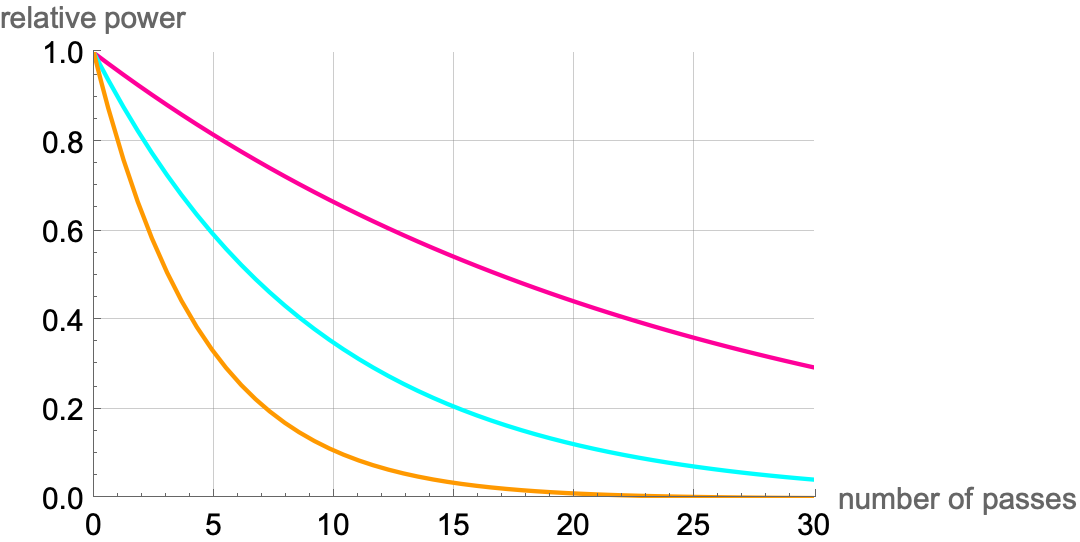}
    \caption{X-ray laser intracavity power loss as a function of the number of cavity passes for 4\% (magenta), 10\% (cyan) and 25 \% (orange) loss per pass. \label{fig:cavity losses}}    
\end{figure}

\begin{table}[h]
    \centering
    \begin{tabular}{|c|c|c|c|c|c|}
    \hline
    Diamond & Bragg angle (deg) & $\Delta \theta$ ($\mu$rad) & $\Delta \omega$ (meV) & $\Delta \omega /\omega \times 10^{-6}$ & $\tau$ (fs)\\
    \hline
    311 & 45.7 & 9.6 & 75.4 & 9.4 & 23.9 \\
    \hline
    400 & 59.8 & 14 & 65.6 & 8.1 & 27.4 \\
    \hline
    331 & 70.3 & 13 & 37.4 & 4.7 & 48.1 \\
    \hline
    \end{tabular}
    \caption{Diamond crystal reflections at 8.048 keV: Miller indices, corresponding Bragg angle and Darwin width $\Delta \theta$ and $\Delta \omega$ in $\mu$rad and meV respectively, relative bandwidth and corresponding transform-limited time duration.}
    \label{tab:crystals}
\end{table}
For a bi-focal cavity, with $f=\L_C / 4.0$, we use one CRL at point B with 2 lenses for a total of 60 micron of Be, resulting in loss of 1.15\% per turn in the CRLs. If we want $f = 7.5$ m we need $R/N=2f\delta=7.7\cdot10^{-5}$ m. For $N=2$ we have $R\approx150$ $\mu$m; a typical wall thickness of $R=150$ $\mu$m CRL is 30 $\mu$m. 

If we assume a transmission per turn to be $T$ then in 9 more turns the total minimum acceptable transmission can be calculated from $T^{9}=0.1$ or $T_{min}=77.4\%$. 
In Figure \ref{fig:cavity losses}, we show the decrease in the intensity of a pulse circulating without amplification for three cases of 4\% , 10\% and 25\% losses per pass.
The case we just discussed of a single CRL lens corresponds to about 10\%.
Considering this case,  Figure \ref{fig:cavity losses} shows that reducing the cavity length from 299.8 m to 29.98 m, the case considered above, we reduce the number of photons acting as seed at the next amplification event by a factor of 0.3.  A more detailed evaluation of the cavity losses will also require calculating the effects of misalignments and is left to future optimization studies. 
We note that if the losses are larger the oscillator would still work using a longer Bragg cavity.
 
 For a bi-focal cavity length of $L \approx 30$ m the distance the x-rays travel before arriving at the CRL lens is about 15 m. If we use a collimator to limit the angular divergence to about $\Delta \theta = 14$ $\mu$rad the photon beam size at the CRL is about 0.22 mm. 
Alternative focusing systems are possible. For instance a glancing incidence elliptical lens is an interesting option to be studied.

\subsection{Projected performance}
\begin{table}
    \centering
    \begin{tabular}{|c|c|}
\hline
Photon energy, keV & 8.048 \\
\hline
Number of photons/pulse & $3\cdot10^8$ \\
\hline
Pulse repetition rate, kHz&125\\
\hline
Average laser power, mW&48\\ 
\hline
Angular distribution width, $\mu$rad & 14\\
\hline
Energy width*, meV & 65.6 \\
\hline
Pulse duration*, fs & 27.4 \\
\hline
 \end{tabular}
    \caption{ XLO-II  characteristics operating at 8.048 keV,  lasing on copper $K\alpha_1$ line.
    * For a given choice of diamond (4,0,0) reflection in the cavity.}
    \label{tab:XLO-II}
\end{table}
If we had no limitation on the LCLS-II HE current, and thus we could operate it at 1 MHz, we could operate XLO-II using four pump pulses to reach saturation, out-couple the pulse, and restart with the next pump pulse. In this case, the out-coupled 8.048 keV pulses repetition rate would be 250 kHz. We assume that we block the following 4 pump pulses after XLO-II has reached saturation via the programmable LCLS-II injector laser shutter.
Operating the pump pulse with an energy of 40 $\mu$J on target, the average LCLS-II HE 9 keV XFEL pulse power is 100 W, below the available limit \cite{Raubenheimer:FLS2018-MOP1WA02}. In this regime, 100 W would arrive at the KB mirror system. A new nano-focus instrument is being built to operate with LCLS-II HE at a high repetition rate. The system will have a zoom focus and will have smaller losses than the one used until now on LCLS, and that we have used in our experimental work on XLO, with 60\% losses. It will be able to operate at an average x-ray pulse power of 20 W \cite{Meng}. The losses in the KB system are due to misalignment, diffraction and absorption. Assuming that 50 W is transmitted, and 50 W of power is lost at the KB mirror system, 10 W will be deposited in the mirrors as heat, while
the limit for the incident XFEL power thermal load on the KB focusing system is about 20 W. However, to respect all other possible limitations we reduce the number of XLO pulses by a factor of 2, using only 50\% of the XFEL pump pulses, with trains of four 40 $\mu$J pulses, separated by 4 $\mu$s and thus reducing XLO-II repetition rate to 125 kHz. The average power for 8.048 keV photons for XLO-II would then correspond to $\approx 3.75\times10^{13}$ photons/s or 48 mW.
The main characteristics of the XLO-II pulses are summarized in Table \ref{tab:XLO-II}.
It is important to emphasize that the 8.048 keV photon pulses described in Table \ref{tab:XLO-II} are fully coherent and transform-limited as shown in Figure \ref{fig:XLO II pass 1}.

\begin{figure}[h]
    \centering
    \includegraphics[width=0.8\linewidth]{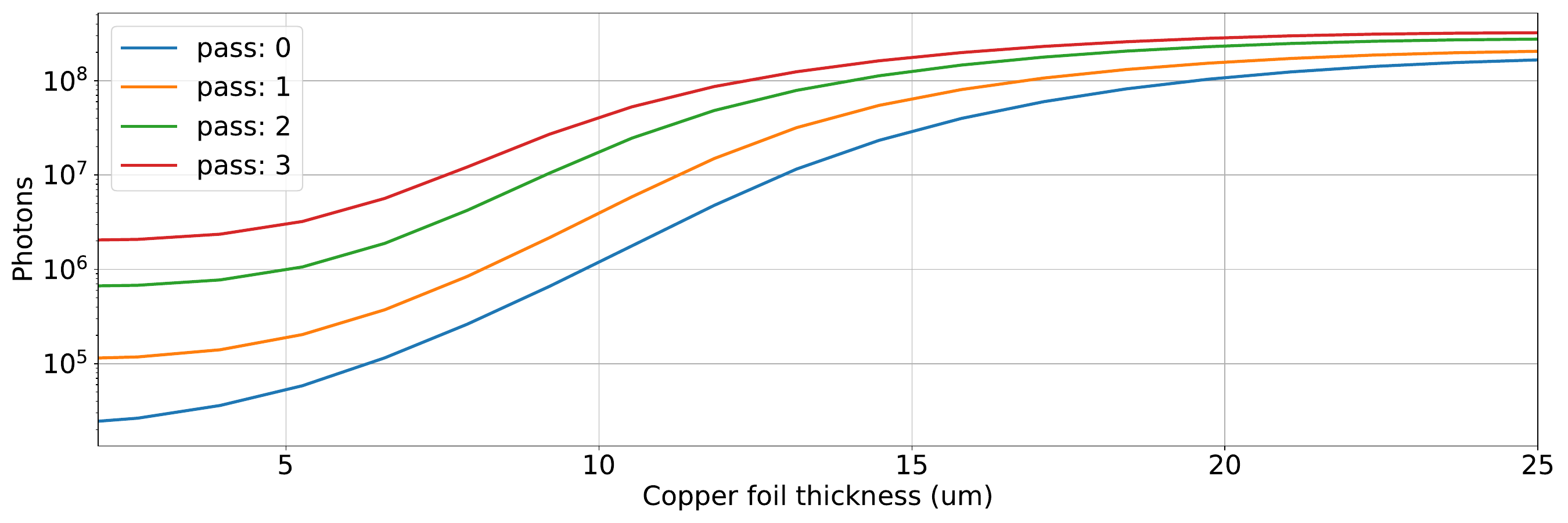}
    \caption{ Number of 8.048 keV photons as a function of gain medium thickness for the case of XLO-II and the first four passes in the cavity.}
    \label{fig:XLO II four passes}
\end{figure}

\begin{figure}[h!]
    \centering
    \includegraphics[width=1\linewidth]{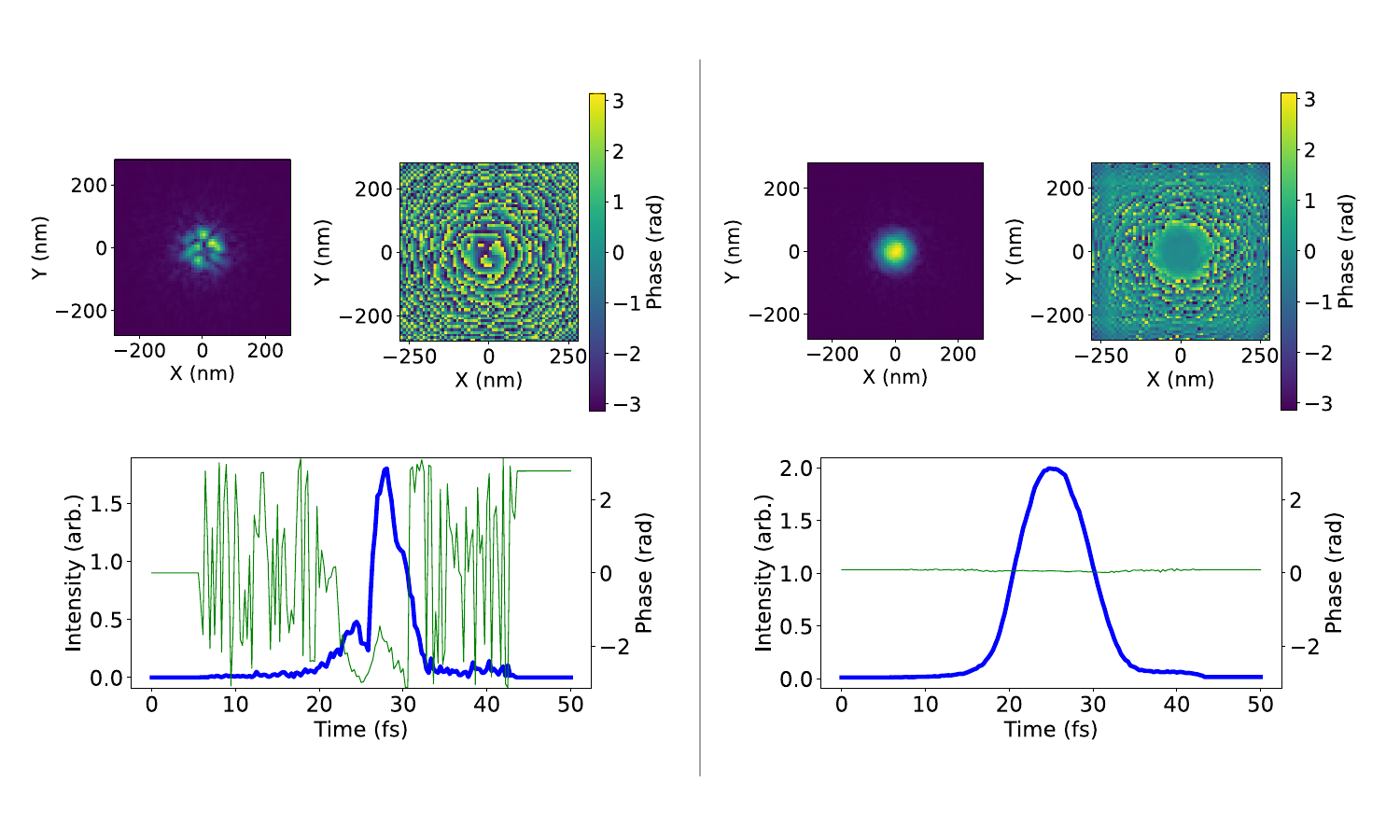}

    \caption{Temporal and spatial pulse profiles after the first pass in the cavity (left panel) and final pulse profiles before out-coupling (right panel).
    The blue and green curves are the pulse amplitude and phase. Notice that after four passes, the phase is constant throughout the pulse, indicating a temporal transform-limited pulse. The relative line width at FWHM is $1.25\times10{^{-5}}$.}
    \label{fig:XLO II pass 1}
\end{figure}

Numerical simulation of the lasing and pulse evolution in the diamond (4,0,0) bow-tie cavity have been done using a state-of-the-art numerical 3D code described in Ref. \cite{benediktovitch2023stochastic} \footnote{We note that only $K\alpha_1$ and $K\alpha_2$ emission lines are presently included in the code. The fine structure of the multiplet lines of $K\alpha_1$ is presently not captured, resulting in slightly narrower simulated ASE bandwidth. }. The results on the number of photons as a function of gain medium thickness are shown in Figure \ref{fig:XLO II four passes}. The simulation reported here is done assuming that the pump pulses are focused to about $10^5$ $J/cm^2$, with an FWHM duration of 10 fs. For these simulations, we assume a presence of small undulator seed, and 30 m cavity length. We consider four passes in the cavity with amplification taking place at each tenth pass. In the initial pass, we start with the undulator-generated seed and reach about $10^8$ photons, in agreement with the results discussed in Section \ref{recent-results} for the case of assumed pump power density. A large fraction of these photons is lost in the first cavity pass because of the selection in angle and energy done by the Bragg mirrors. 
About 1 in a thousand survive to act as a seed for the second amplification pass; this will be explained below. After the second amplification pass the cavity losses are much smaller and the oscillator reaches saturation in as few as four passes with about $3\cdot10^8$ photons per pulse.
The shape of the 8.048 keV photon pulse after one and four cavity pass with amplification is shown in Figure \ref{fig:XLO II pass 1}. We can notice that in four passes the output pulse becomes essentially transform limited with a constant phase throughout the pulse.

 The numerical simulations presented here have been made for a pump pulse energy, power density and a population inversion volume that is about equal to those used for the measurements shown in Figure \ref{fig:XLO II gain curves}.
The experimentally obtained number of approximately $10^8$ stimulated emission $K\alpha_1$ photons per pulse in the case undulator seeding supports the validity of the 3D model used in the simulations provided here.

It is important to note that the pulse profile shown in Figure \ref{fig:XLO II pass 1} is evaluated at the exit from the gain medium and depends on the pump pulse and Bragg cavity parameters. If the pulse does more Bragg reflections in the cavity, its bandwidth and consequently pulse length will be further modified. In our example, the pulse duration after the fourth pass is approximately 12 fs FWHM. The corresponding energy bandwidth for a transform-limited pulse, obtained from the relationship $\Delta E$ $\times$ $\tau$ = 1.8 eV-fs, results in $\Delta E= 0.15$ eV FWHM. Different pump pulse length, can result in different transform limited XLO energy resolutions/pulse lengths. Because the energy resolution corresponding to the Darwin width of a diamond crystal Bragg reflection in backscattering is smaller than 0.15 eV, the XLO pulse will be stretched if it passes through one or more Bragg reflections before exiting the cavity. For example, using the values of diamond crystal Bragg reflections given in Tab. \ref{tab:crystals}, the (4,0,0) reflection has a bandwidth of $\Delta E$= 0.065 eV at 8.048 keV, corresponding to a pulse length of $\tau$=27 fs. This shows that the XLO bandwidth/pulse length depends on whether the out-coupling occurs directly after the gain medium or after additional Bragg reflection(s). Depending on the experimental requirements, researchers can operate XLO-II with various out-coupling geometries resulting in different bandwidths.   

The limit to the number of photon/pulse is related to the volume where we generate the population inversion. By increasing the volume while maintaining constant pump pulse power density, we would proportionally increase the number of excited atoms and the number of output photons. How to best accomplish this will be the subject of future optimization studies.

\subsection{Gain medium of XLO-II}
\begin{figure}[h]
    \centering
    \includegraphics[width=0.5\linewidth]{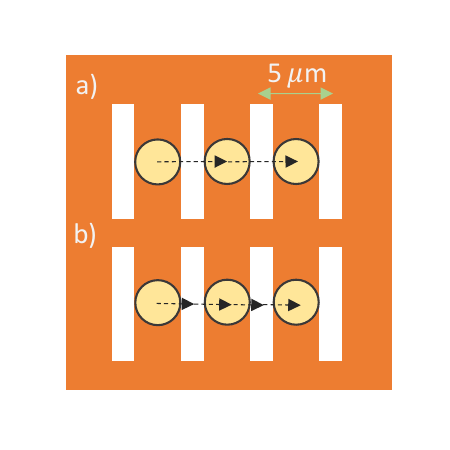}
    \caption{A section of a XLO-II copper target with empty spaces (shown in white). The target is scanned first horizontally. After completing one row the target is stepped to the next one. As the x-ray pulses recirculate in the cavity and the delay between pump pulses is 1 $\mu$s, moving by 5 $\mu$m in 1 $\mu$s requires a speed of 5 m/s (case a)), provided by a stepping motor, or 25 m/s in a jump-rest-jump regime (case b)). The thickness of the copper stripe is selected in accordance with Fig. \ref{fig:XLO II copper damage}.}
    \label{fig:gain medium design }
\end{figure}

As previously discussed \cite{Alex2}, the most effective gain medium is a pure copper metal foil with 25 $\mu$m thickness. The operation of XLO-II, where the cavity round trip time is much shorter than the time interval between pump pulses, requires a new design that avoids absorption of the circulating pulse by the gain medium during round trips where there is no pump pulse. Instead of a continuously rotating wheel \cite{10.1063/5.0168125}, we propose to employ a micro-machined copper sheet with perforated stripes. Copper alternates with transparent slots where the circulating seed pulse goes through. 
The sketch is shown in Figure \ref{fig:gain medium design }.
For a 30 m long cavity we can envision that the horizontal periodicity is 5 $\mu$m and the empty space between copper stripes is 1 $\mu$m. Depending on the results of future gain medium R\&D, we foresee two modes of gain medium operation. One option is that the initial hole generated by the pump pulse removes most of the copper between the transparent slots. In that case the foil can be moved at a constant velocity to reach the next copper section at the time when the next pump pulse arrives. For a center-to-center pitch of 5 $\mu$m and 1 MHz pump pulse repetition rate, this corresponds to a constant velocity of 5 m/s. An alternate option of operating the gain medium is by moving it by 2.5 
$\mu$m to the center of the empty space after the first pump pulse before the circulating x-ray pulse. At 30 m cavity length, this move needs to be complete after 100 ns = 0.1 $\mu$s requiring a velocity of 2.5 $\mu$m/0.1 $\mu$s = 25 m/s. There the sheet stays for 0.8 $\mu$s, and in the subsequent 0.1 $\mu$s it moves to the center of the copper area, again at 25 m/s in time for the next pump pulse to arrive. The slots can be organized in a planar structure with a vertical separation of choice. This structure is traversed by the pump pules during the oscillator operation.

Both, the constant velocity and the jump-rest-jump operations, should be achievable with existing mechanical components. The parameters discussed here are an example based on our numerical simulations, and the choice and design. Depending on the cavity length and number of pulses needed to reach saturation and out-couple the XLO pulses, the exact gain medium parameters might deviate from these numbers. Future R\&D will explore this and other potential configurations to optimize the XLO-II gain medium system. For instance, instead of slots it is also possible to use circular holes with the diameter equal to the slot width. The best practical solution will be studied and tested in future experiments.

\section{Conclusions}
We have discussed the feasibility of a high-repetition rate x-ray laser oscillator, XLO-II, using recent experimental results and numerical analysis. XLO-II will generate intense, highly monochromatic ($\Delta E$/E $\approx$ $10{^{-5}}$), transform-limited 6-10 keV x-ray pulses at repetition rates up to 125 kHz when using SASE pulses for the gain medium population inversion and initial seed from a superconducting 1 MHz repetition-rate XFEL, like LCLS-II-HE. XLO-II will be a unique new x-ray laser source for research areas including x-ray interferometry, x-ray quantum optics, and x-ray coherent imaging. As was the case for laser oscillators, once made available, XLO-II might find other applications not yet envisioned.

\section{Acknowledgements}
This work is supported by the U.S. Department of Energy Contract No. DE-AC02-76SF00515. The use of LCLS and SSRL is supported by US DOE, Office of Science, OBES (contract No. DE-AC02-76SF00515), the work at SACLA was performed with the approval of the Japan Synchrotron Radiation Research Institute (proposal no. 2017B8066).

We wish to thank Mengning Liang, Tor Raubenheimer, Andrew Aquila, Diling Zhu, Thomas Kroll, Christopher J. Takacs (SLAC) for many useful discussions and information; Ichiro Inoue (SACLA), Taito Osaka (SPring-8/SACLA), Jumpei Yamada (SACLA/Osaka University) for their consultation on hard x-ray optics; Kenan Li and Anne Sakdinawat (SLAC) for their help with scanning electron microscope analysis; Thomas Linker (SLAC), Stasis Chuchurka and Nina Rohringer (DESY) for help with the lasing theory and simulations; Pratik Manwani (UCLA) and Eric Galtier (SLAC) for gain medium damage simulation work; Margaret Doyle (University of California - Berkeley), River Robles (SLAC), Nathan Majernik (University of California - Los-Angeles/SLAC) and Noah Welke (University of Wisconsin - Madison/Northwestern University) for their help with the experiments and data analysis.


\end{document}